\documentclass[%
 reprint,
 amsmath,amssymb,
 aps,
]{revtex4-2}

\usepackage{tabularray}
\usepackage{graphicx}
\usepackage{dcolumn}
\usepackage{bm}
\usepackage{upgreek}
\usepackage[english]{babel}
\usepackage{hyperref}
\hypersetup{
    colorlinks=true,
    linkcolor=blue,
    citecolor=blue,
    urlcolor=black
}

\begin{document}

\preprint{APS/123-QED}

\title{\textbf{Open-Source Highly Parallel Electromagnetic Simulations for Superconducting Circuits} 
}%

\author{David Sommers}
\author{Zach Degnan}
\author{Divita Gautam}
\author{Yi-Hsun Chen}
\author{Chun-Ching Chiu}
\author{Arkady Fedorov}
\author{Prasanna Pakkiam}
\email{Contact author: p.pakkiam@uq.edu.au}
\affiliation{School of Mathematics and Physics, University of Queensland, Brisbane, Australia\\
}

\date{\today}

\begin{abstract}
Electromagnetic simulations form an indispensable part of the design and optimization process for superconducting quantum devices. Although several commercial platforms exist, open-source alternatives optimized for high-performance computing remain limited. To address this gap, we introduce SQDMetal, a Python-based API that integrates Qiskit Metal (IBM), Gmsh, Palace (AWS), and Paraview (Kitware) into an open-source, highly parallel simulation workflow for superconducting quantum circuits. SQDMetal enables accurate, efficient, and scalable simulations while remaining community-driven and free from commercial constraints. In this work, we validate SQDMetal through mesh convergence studies which benchmark SQDMetal against COMSOL Multiphysics and Ansys, demonstrating excellent agreement for both eigenmode and electrostatic (capacitance) simulations. Furthermore, we simulate superconducting resonators and transmon qubits, showing reasonable agreement with experimental measurements. SQDMetal also supports advanced capabilities, including Hamiltonian extraction via the energy participation ratio (EPR) method, incorporation of kinetic inductance effects, and full 3D modeling of device geometry for improved predictive accuracy. By unifying open-source tools into a single framework, SQDMetal lowers the barriers to entry for community members seeking to access high-performance simulations to assist in the design and optimization of their devices.

\end{abstract}

\maketitle


\section{Introduction}

The field of superconducting quantum circuits has emerged as one of the leading theoretical and experimental platforms in the pursuit of fault-tolerant quantum computing \cite{arute_quantum_2019, wu_strong_2021}. Over the past two decades, this maturing field has experienced significant growth, highlighted by the evolving sophistication of experimental techniques and circuit design to improve qubit coherence times \cite{kjaergaard_superconducting_2020}. As the field continues to develop, focus is shifting from understanding the physics underlying superconducting circuit behaviour \cite{nobel} to addressing engineering challenges, such as the design and optimization of device geometries \cite{levenson-falk_review_2024}. Computational tools such as electromagnetic simulation software have become increasingly important for optimizing superconducting circuit design \cite{shammah_open_2024, levenson-falk_review_2024}. Numerous frameworks have been proposed to simulate device behaviour and extract the circuit Hamiltonian using electromagnetic solvers. Among the most widely adopted are the lumped oscillator model (LOM) \cite{minev_circuit_2021, chitta_computer-aided_2022}, blackbox quantization (BBQ) \cite{nigg_black-box_2012}, and the energy participation ratio (EPR) \cite{minev_energy-participation_2021}, which employ electrostatic, driven frequency domain, and eigenmode simulations, respectively. These approaches enable designers to evaluate, refine, and optimize circuit parameters prior to fabrication. However, performing such simulations remains challenging for researchers and engineers with limited access to proprietary electromagnetic solvers such as Ansys HFSS or COMSOL.\\ 

Furthermore, implementing commercial electromagnetic solvers on high-performance computing (HPC) systems to achieve greater simulation efficiency and accuracy often incurs significant licensing costs. In this work, we present an accurate, efficient, and fully open-source simulation workflow that leverages high-performance parallelization to remove the dependence on commercial software, thereby lowering barriers to entry for research groups and fostering a community-driven approach.
The simulation package, which we have termed SQDMetal, is a Python application programming interface (API) that integrates existing open-source tools for the design and simulation of superconducting quantum devices. The core software components of SQDMetal include:

\begin{itemize}
\item \textbf{Qiskit Metal} \cite{QiskitMetal} — an open-source framework developed by IBM for the design of superconducting quantum chips and devices.
\item \textbf{Gmsh} \cite{geuzaine_gmsh_2009} — an open-source finite element mesh generator developed by Christophe Geuzaine and Jean-François Remacle for creating 3D meshes and models.
\item \textbf{Palace} (Parallel Large-scale Computational Electromagnetics) \cite{palace} — an open-source electromagnetic solver developed by Amazon Web Services (AWS) for HPC environments, enabling faster and more accurate large-scale simulations.
\item \textbf{ParaView} \cite{paraview} — an open-source visualization and post-processing tool developed by Kitware for analyzing and rendering finite element simulation results.
\end{itemize}
Our major contribution in this work is the integration of these open-source software packages into a cohesive workflow, enabling design and simulation to be executed from within a single Jupyter notebook. Our framework provides control over meshing and simulation parameters through a set of simple commands. We also plan to expand the functionality of SQDMetal to address evolving community needs. Our goal is to promote usability and encourage programmatic design which enhances code re-usability. Through this integration, we aim to deliver an open-source platform that the community can readily adopt and extend.

This paper presents SQDMetal’s architecture, validation, and applications in superconducting circuit simulation. Section II describes the device design and simulation workflow. Section III compares the SQDMetal API with existing commercial solvers (Ansys and COMSOL Multiphysics) to validate its accuracy and reliability via a mesh convergence study. Finally, Section IV presents a comparison between simulation results obtained with SQDMetal and experimental measurements of superconducting resonators and qubits, showing good agreement in the process.

\begin{figure*}[!]
\includegraphics[width=1\textwidth]{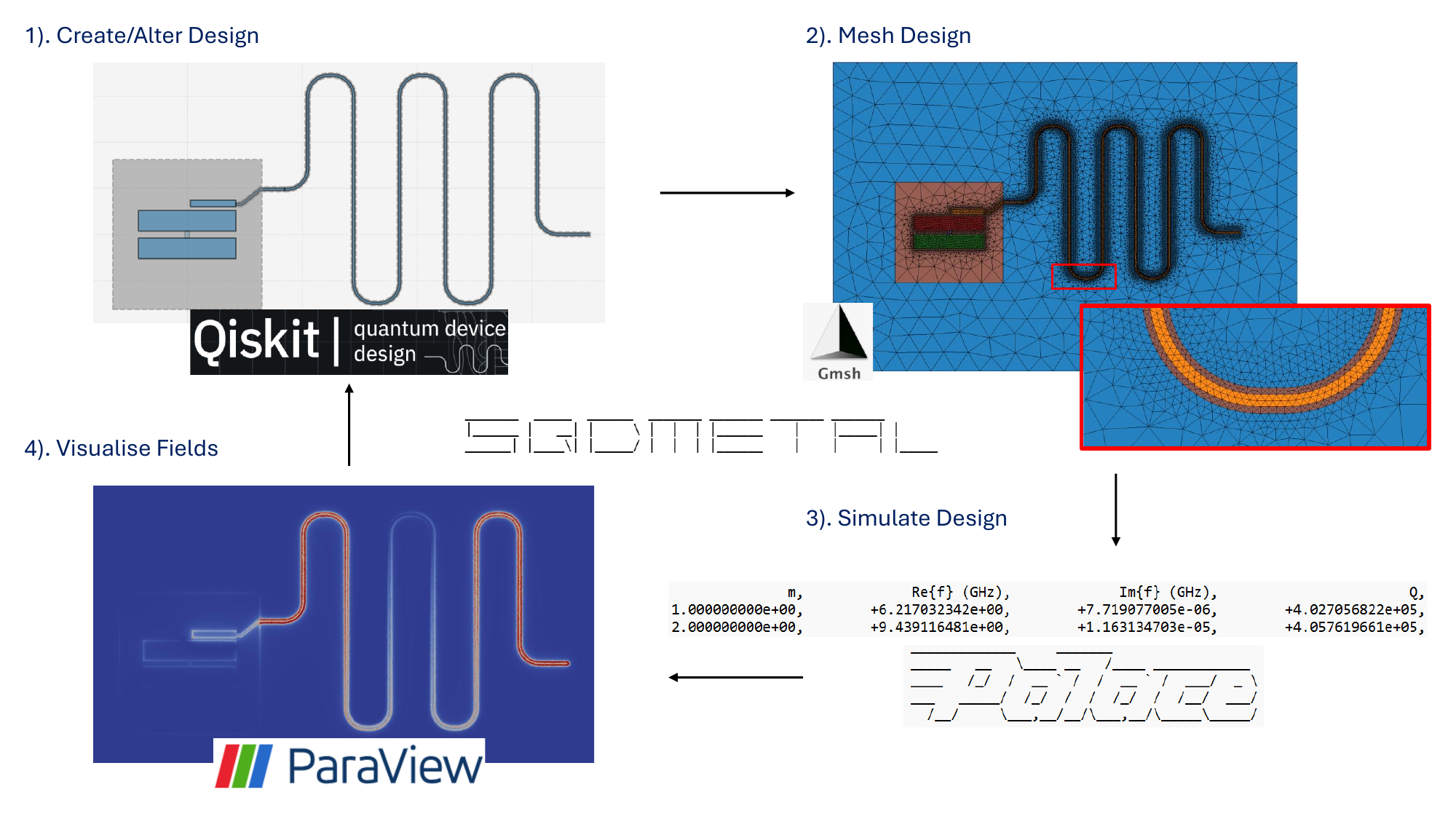}
\caption{\label{fig:workflow} Visualization of the SQDMetal workflow and the software packages used for each process.}
\end{figure*}

\section{Workflow}

The operation of the SQDMetal API is shown in Fig.~\ref{fig:workflow}, providing a diagrammatic representation of our workflow. The process begins with Qiskit Metal, where the superconducting device is designed. Qiskit Metal features an extensive component library and enables users to programmatically develop circuit layouts in Python via Jupyter notebooks \cite{QiskitMetal}. Qiskit Metal also supports community-driven development, allowing users to create and share custom components that can be integrated into new designs~\cite{shanto_squadds_2024}.

Once the initial device design is complete, our API reconstructs the geometry in Gmsh and generates a finite-element mesh according to user-defined parameters. Users can control the meshing process through simple commands that specify the minimum and maximum element sizes and the distance over which transitions occur. SQDMetal also incorporates the adaptive mesh refinement (AMR) feature available in Palace, allowing users to define a seed mesh that is automatically refined by the solver to improve simulation accuracy. Because the quality of the finite-element mesh directly affects the accuracy of electromagnetic simulations, SQDMetal provides fine-grained control to ensure that device geometries are meshed with sufficient resolution. 

After meshing, the SQDMetal API transfers the generated mesh file to Palace to perform finite-element electromagnetic simulations. Palace is developed on top of MFEM, an open-source high-performance C++ library for finite-element discretization, and is designed to operate across a wide range of platforms from laptops to supercomputers~\cite{anderson_mfem_2021}. While proprietary software packages such as COMSOL Multiphysics and Ansys HFSS also support finite-element electromagnetic simulations on HPCs, there are few open-source alternatives. Palace addresses this gap by providing scalable, open-source computational electromagnetics capabilities. Within SQDMetal, we have integrated Palace’s functionality to perform the following types of simulations:

\begin{enumerate}
\item Electrostatic simulations to compute capacitance matrices,
\item Eigenmode simulations to determine eigenfrequencies, quality factors, and energy participation ratios (EPR), and
\item Driven radio-frequency (RF) simulations to extract scattering parameters.
\end{enumerate}

In the final step of the workflow, SQDMetal utilizes ParaView to visualize the results of the simulation, including the computed electric and magnetic field distributions. Using Paraview, users can verify the simulated fields exhibit physically consistent behaviour, ensuring that the simulation accurately represents device performance. ParaView also provides extensive post-processing capabilities, such as numerical integration and quantitative field analysis, allowing users to extract key metrics directly from the simulation data.

At this stage, a full cycle of the SQDMetal workflow has been completed. Users can then modify their device design based on the simulation results and re-run the workflow, establishing an iterative process between design, simulation, and analysis that facilitates efficient device optimization.

\section{Mesh Convergence Study}

To evaluate the accuracy of the SQDMetal simulation workflow, we compare its results with those obtained using two widely employed commercial solvers: COMSOL Multiphysics and Ansys HFSS. This benchmarking study is designed to verify that the accuracy of SQDMetal is comparable to that of established tools currently used in the superconducting quantum device community.

\begin{figure*}[!]
\includegraphics{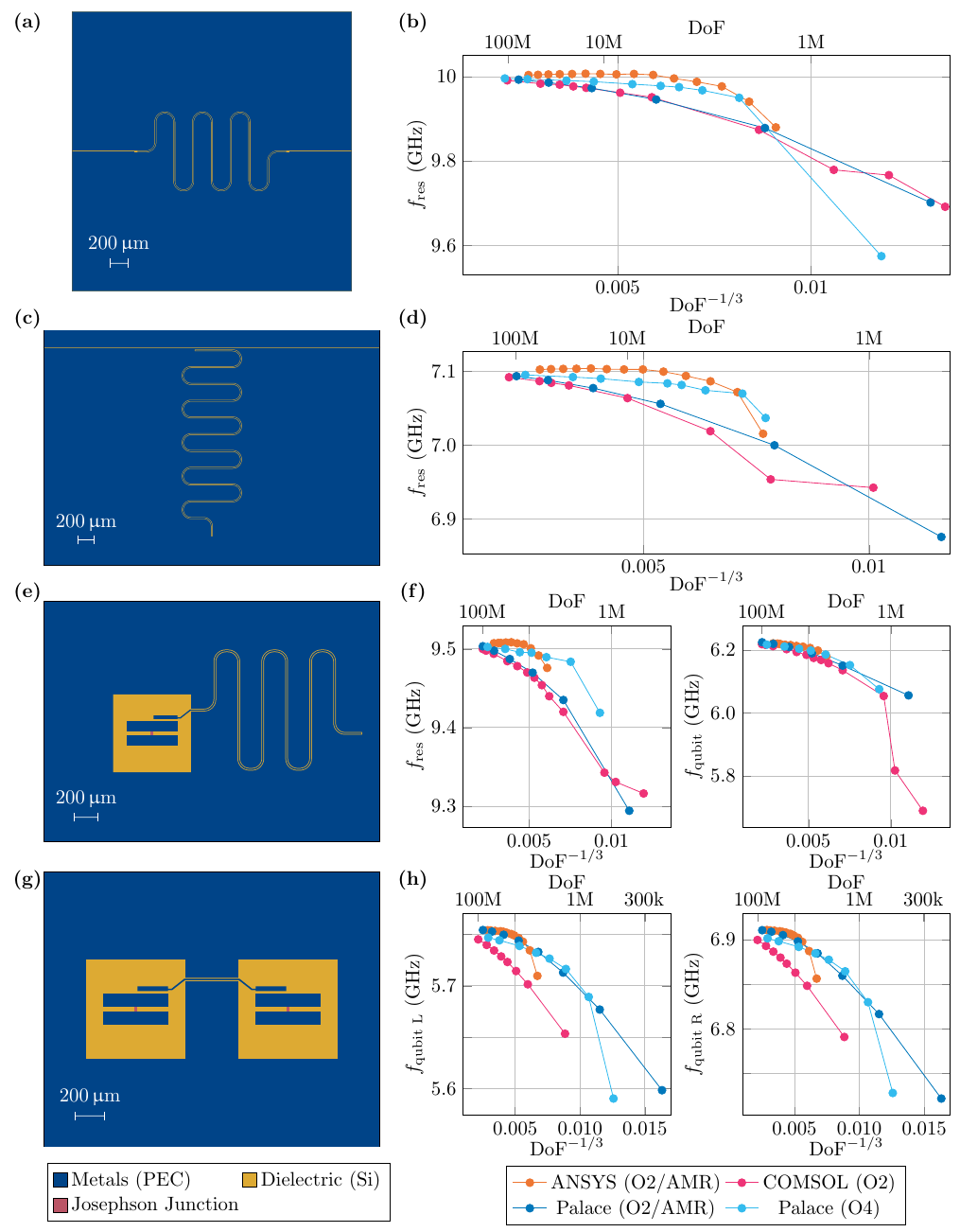}
\caption{\label{fig:eig_results} \textbf{Benchmark designs and associated eigenmode simulations}. The simulations were performed using ANSYS, COMSOL and Palace. In each simulation, the mesh size (${\sim}{\text{DoF}}$) is decreased to converge to the continuum limit toward the LHS of the plots where $\text{DoF}\to\infty$. \textbf{(a)}-\textbf{(b)} A single CPW resonator with its mode $f_{\text{res}}$. \textbf{(c)}-\textbf{(d)} A CPW resonator with mode $f_{\text{res}}$ capacitively coupled to a feedline. \textbf{(e)}-\textbf{(f)} Transmon coupled to a CPW resonator with modes $f_{\text{qubit}}$ and $f_{\text{res}}$ respectively. \textbf{(g)}-\textbf{(h)} Two capacitively coupled transmons with modes $f_{\text{qubit L}}$ and $f_{\text{qubit R}}$ for the left and right qubits respectively. In all simulations $f_{\text{res}}$ refers to the fundamental resonator mode while $f_{\text{qubit}}$ refers to the qubit mode}
\end{figure*}

In the first part of this study, four device geometries, depicted in Fig.~\ref{fig:eig_results}, are analyzed using eigenmode simulations. The selected designs were:

\begin{enumerate}
\item a single coplanar waveguide (CPW) resonator,
\item a CPW resonator capacitively coupled to a feedline,
\item a transmon capacitively coupled to a CPW resonator, and
\item two transmons capacitively coupled to each other,
\end{enumerate}
where the last two designs were taken from the Qiskit Metal tutorial repository. All geometric and material parameters for each device were identical across solvers to ensure a fair comparison. The first two designs were defined on a $500 \upmu\text{m}$-thick silicon substrate, while the third and fourth designs, used a $750 \upmu\text{m}$-thick substrate. In all cases, the silicon substrate was assigned a relative permittivity of 11.45, a value commonly used for silicon at cryogenic temperatures~\cite{krupka_measurements_2006}. The superconducting regions were modeled as zero-thickness surfaces with perfect electric conductor (PEC) boundary conditions. Each chip was enclosed in a rectangular box whose interior volume was defined as vacuum, with all bounding surfaces set to PEC to emulate placement within a metallic sample holder. The enclosing box was twice the height of the chip, while its width and length were 20\% larger than the chip. The substrate was placed on the floor in the centre of the enclosing box, thereby grounding the bottom of the device. For the first two designs, the ends of the feedline were terminated with 50 $\Omega$ lumped ports to prevent signal reflections and ensure well-defined boundary conditions.

To compare the various simulation platforms, we performed a mesh-convergence study in which we start with a coarse mesh and progressively refine the mesh until convergence of the eigenmode frequency is observed. In general, to increase the accuracy of a finite element simulation, two properties of the mesh can be changed:

\begin{enumerate}
    \item the size of the mesh elements, and
    \item the polynomial order of the basis functions. 
\end{enumerate}
Decreasing the size of mesh elements allows vector fields (e.g. the electric field in an eigenmode simulation) and scalar fields (e.g., the electric potential in an electrostatic simulation) to be captured with higher spatial resolution, resulting in more accurate solutions. For this benchmark study, we employ exclusively tetrahedral mesh elements. Within each element, the field quantities are interpolated using basis (shape) functions that approximate the field variation across the element~\cite{polycarpou_introduction_2006, jin_theory_2010}. Higher-order basis functions improve the accuracy of field interpolation but increase computational cost (in both memory and time). The total number of global unknowns that an electromagnetic solver must compute are referred to as the degrees of freedom (DoF). The DoF directly determine the size of the global matrix system that is constructed at the beginning of the simulation and thus is an indicator of the computational resources required. Specifically, the size of the system matrices for eigenmode or electrostatic simulations is $n \times n$, where $n$ is the DoF. Increasing the number of mesh elements and order of the basis functions both increase the DoF.
 
To compare simulation platforms, two methods were employed to increase the number of DoF:

\begin{enumerate}
\item Manual mesh refinement (MMR), and
\item Adaptive mesh refinement (AMR).
\end{enumerate}
When implementing manual mesh refinement (MMR), the user increases the DoF by visually inspecting the electric field distribution and locally decreasing the mesh element size, that is, increasing the mesh element density in regions of strong electric field magnitude. SQDMetal provides specific commands that allow users to selectively refine regions of a device or individual circuit components. However, achieving convergence of a given parameter through MMR can be time-consuming and requires considerable user intuition. The alternative approach, adaptive mesh refinement (AMR), requires less user input while achieving excellent accuracy. In this method, the solver automatically refines the mesh by computing an error indicator that estimates the local deviation between the numerically computed electric field within an element and a smoothed field value calculated in post-processing~\cite{zienkiewicz_simple_1987,nicaise_zienkiewiczzhu_2005}. Elements exceeding a user-defined error threshold are subdivided into smaller elements until the estimated error falls below that threshold. AMR requires a reasonably well-refined initial seed mesh to be defined by the user in order to produce the most accurate results \cite{levenson-falk_review_2024}. For both refinement schemes in our study, the polynomial order of the basis functions is held constant; therefore, the convergence behaviour observed in this study arises solely from changes in the mesh element size within regions of high electric field. 

We emphasise that the objective of this mesh convergence study is not to determine which solver has the fastest convergence rate or simulation speed, but rather to demonstrate that each solver exhibits agreement on the simulated values as the DoF increases towards the continuum limit.

\subsection{Eigenmode Convergence Results}

The results of the eigenmode mesh convergence study for each design are presented in Fig.~\ref{fig:eig_results}. Each panel in the figure shows a schematic of the device geometry alongside its corresponding plot of resonant frequency as the DoF are increased. In the schematics, blue regions depict superconducting metals, modelled as PEC, while gold regions indicate the exposed silicon substrate. The red rectangles between the capacitor pads are Josephson junctions modelled as lumped element ports. Each point in the plots represents a single eigenmode simulation conducted at a given DoF.

The data points in Fig.~\ref{fig:eig_results} correspond to results from different solvers and mesh refinement strategies. Results for Palace simulations using AMR and second order basis functions (AMR, O2) are shown as dark blue circles, while simulations conducted using Palace and MMR with fourth order basis functions (MMR, O4) are shown as light blue circles. Ansys HFSS simulations using AMR and second order basis functions (AMR, O2) are plotted in orange, while COMSOL Multiphysics simulations using MMR and second order basis functions (MMR, O2) are in red. For this study, we define a notional mesh element size as $m = \text{DoF}^{-1/3}$~\cite{palace_tutorial} where $m\to0$ as $\text{DoF}\to\infty$ to approach the continuum limit. The simulated resonant frequencies are plotted against this quantity to illustrate convergence of each solver toward a common frequency value. The corresponding total DoF for each simulation is indicated along the top axis of each plot for reference.

\begin{table}[h!] 
\caption{ \label{tab: eig_results} \textbf{Final eigenmode simulation results}. Comparison of the final simulated resonance frequencies for each device design and solver (Palace, COMSOL, and Ansys HFSS) obtained from the eigenmode convergence study shown in shown in Fig. \ref{fig:eig_results}.}
\begin{ruledtabular}
\begin{tabular}{lccc}
Solver & DoF & Mode 1 (GHz) & Mode 2 (GHz)  \\
\hline
\multicolumn{4}{c}{Single Resonator}\\
\hline
Palace MMR    & 111,374,732   & 9.996   & - \\
Palace AMR    & 69,301,924   & 9.993    & - \\
Comsol    & 102,109,058   & 9.991   & -   \\
Ansys     & 51,411,402  & 10.004   & -   \\ 
\hline
\multicolumn{4}{c}{Resonator Coupled to Feedline}\\
\hline
Palace MMR    & 74,348,288   & 7.095  & -  \\
Palace AMR    & 96,313,892   & 7.094  & -   \\
Comsol     & 122,062,996  & 7.092  & -   \\
Ansys      & 50,743,641   & 7.101  & -    \\
\hline
\multicolumn{4}{c}{Transmon Coupled to Resonator}\\
\hline
Palace MMR    & 65,688,988  &  6.217  & 9.503  \\
Palace AMR  & 95,143,250   & 6.225  & 9.507   \\
Comsol     & 96,184,134   & 6.218  & 9.500   \\
Ansys      & 42,708,384   & 6.221  & 9.507    \\
\hline
\multicolumn{4}{c}{Two Coupled Transmons}\\
\hline
Palace MMR    & 39,927,764  & 5.747  &  6.902  \\
Palace AMR    & 62,522,530   & 5.754  & 6.911   \\
Comsol     & 99,657,414   & 5.745  & 6.900   \\
Ansys    & 41,449,212   & 5.754  & 6.911  
\end{tabular}
\end{ruledtabular}
\end{table}

The first benchmark design, shown in Fig.~\ref{fig:eig_results}a, is a single coplanar waveguide (CPW) resonator. This geometry was selected as a baseline case, since a $\lambda/2$ resonator is a common component in superconducting circuits. The corresponding plot in Fig.~\ref{fig:eig_results}b shows the convergence of the simulated resonant frequency. The resonator was designed to operate at 10 GHz based on the analytical expression for a half-wavelength CPW resonator~\cite{goppl_coplanar_2008}:

\begin{align}
    f = \frac{c}{2\ell\sqrt{\varepsilon_{\text{eff}}}}, \label{eq:freq}
\end{align}
where $f$ is the resonant frequency of the fundamental mode, $c$ is the speed of light in vacuum,  $\ell$ is the length of the resonator and $\varepsilon_{\text{eff}}$ is the effective permittivity that accounts for field penetration into both vacuum and the substrate. The effective permittivity was approximated as $\varepsilon_{\text{eff}} = (\varepsilon_{\text{Si}} + \varepsilon_{\text{vac}})/2$, where $\varepsilon_{\text{Si}} = 11.45$ \cite{krupka_measurements_2006} and $\varepsilon_{\text{vac}} = 1$. Substituting $f=10$ GHz yields a resonator length of 6.012 mm.

As shown in Fig.~\ref{fig:eig_results}b, all solvers exhibit convergence toward a resonant frequency near 10 GHz as the mesh is refined. The final frequency arrived at by Palace (MMR, O4) was 9.996 GHz, while Palace (AMR, O2) produced 9.993 GHz. The COMSOL Multiphysics and Ansys HFSS solvers produced a frequency of 9.991 GHz and 10.004 GHz, respectively. The percentage differences between Palace (AMR, O2) and the commercial solvers were 0.020\% (COMSOL) and 0.110\%  (Ansys), indicating excellent agreement across all platforms. The electric field distribution is shown in Appendix~\ref{fields_vis}. Table~\ref{tab: eig_results} summarizes the final frequencies and associated DoF for each design included in the eigenmode convergence study. For brevity, in the proceeding sections, percentage differences between Palace (AMR, O2) and commercial solvers are reported in parentheses after the converged frequency values.

The second benchmark design, a coplanar waveguide (CPW) resonator capacitively coupled to a feedline, is shown in Fig.~\ref{fig:eig_results}c, with corresponding eigenmode simulation results presented in Fig.~\ref{fig:eig_results}d. This configuration introduces an additional degree of complexity relative to the single-resonator case via a capacitive coupling to a transmission feedline. For this $\lambda$/2 resonator, a length of 8.475 mm was used, which, according to Eq.~\ref{eq:freq}, yields an expected fundamental frequency of approximately 7.089 GHz. As shown in Fig.~\ref{fig:eig_results}d, at the highest DoF, Palace (MMR, O4) arrived at 7.095 GHz, and Palace (AMR, O2) produced 7.094 GHz. COMSOL Multiphysics and Ansys HFSS produced frequencies of 7.092 GHz (0.028\%) and 7.101  GHz(0.097\%), respectively.

The next benchmark design, shown in Fig.~\ref{fig:eig_results}e, consists of a transmon capacitively coupled to a coplanar waveguide (CPW) resonator. This geometry, adapted from the Qiskit Metal tutorial repository, was selected as a representative benchmark due to its simplicity and familiarity within the superconducting device community. The corresponding eigenmode simulation results are presented in Fig.~\ref{fig:eig_results}f. In this configuration, the transmon was assigned a Josephson inductance of 11 nH, and the resonator length was 6 mm (excluding the coupling region). In these simulations, we compute the linearized qubit frequency. This parameter plays a key role in the energy participation ratio (EPR) method used to extract circuit Hamiltonian parameters~\cite{minev_energy-participation_2021}, which is discussed further in Appendix~\ref{EPR_section}. The solvers again exhibit close agreement as the mesh is refined. At the highest DoF, Palace (MMR, O4) arrived at 6.217 GHz for the transmon and 9.503 GHz for the resonator, while Palace (AMR, O2) arrived at 6.225 GHz and 9.507 GHz, respectively. COMSOL yielded 6.218 GHz (0.112\%) and 9.500 GHz (0.074\%), and Ansys HFSS produced 6.221 GHz (0.064\%) and 9.507 GHz (0.000\%) for the qubit and resonator modes, respectively.

The final benchmark design, consisting of two capacitively coupled transmons, is shown in Fig.~\ref{fig:eig_results}g, with the corresponding eigenmode simulation results presented in Fig. \ref{fig:eig_results}h. This configuration, adapted from the Qiskit Metal tutorial repository, was selected because qubit–qubit coupling is a fundamental aspect of circuit quantum electrodynamics (cQED) and quantum computing architectures. In this design, transmon L and transmon R were assigned Josephson inductances of 13 nH and 9 nH, respectively. As in the previous cases, the simulations solve for the linearized transmon frequencies, corresponding to the eigenmodes of the coupled system. At the highest DoF, the Palace (MMR, O4) solver arrived at 5.747 GHz and 6.902 GHz for transmon L and transmon R, respectively, while Palace (AMR, O2) produced 5.754 GHz and 6.911 GHz. COMSOL yielded 5.745 GHz (0.156\%) and 6.900 GHz (0.159\%), while Ansys HFSS produced 5.754 GHz (0.000\%) and 6.911 GHz (0.000\%). Note that we rounded the percentages to three decimal places. 

These results confirm that all solvers converge towards nearly identical eigenfrequencies, demonstrating that Palace achieves parity with commercial FEM tools. As shown in Table~\ref{tab: eig_results}, the total DoF achieved differs significantly between solvers, particularly for the Ansys HFSS simulations. This variation primarily arises from hardware limitations. The Ansys simulations were performed on a local workstation equipped with approximately 768 GB of RAM. See Appendix \ref{app: hardware} for details on the computers/hardware used in this study. During benchmarking, each simulation was allowed to proceed until available memory was exhausted and the solver could no longer store the matrix system after several iterations of mesh refinement. The same procedure was followed for COMSOL, although higher DoF values were reached. This difference likely reflects variations in how each solver internally stores and manages matrix systems, which is information inaccessible to users. In contrast, the Palace simulations were executed on Bunya, the high-performance computing (HPC) cluster at the University of Queensland~\cite{bunya}. Using the HPC allowed us to utilize larger memory resources, enabling us to simulate with substantially higher DoF. This demonstrates the advantage of HPC-based simulation workflows such as SQDMetal, which can exploit scalable computing resources to achieve improved numerical resolution and accuracy.

\subsection{Capacitance Convergence Results}

\begin{figure}[!htb]
\includegraphics{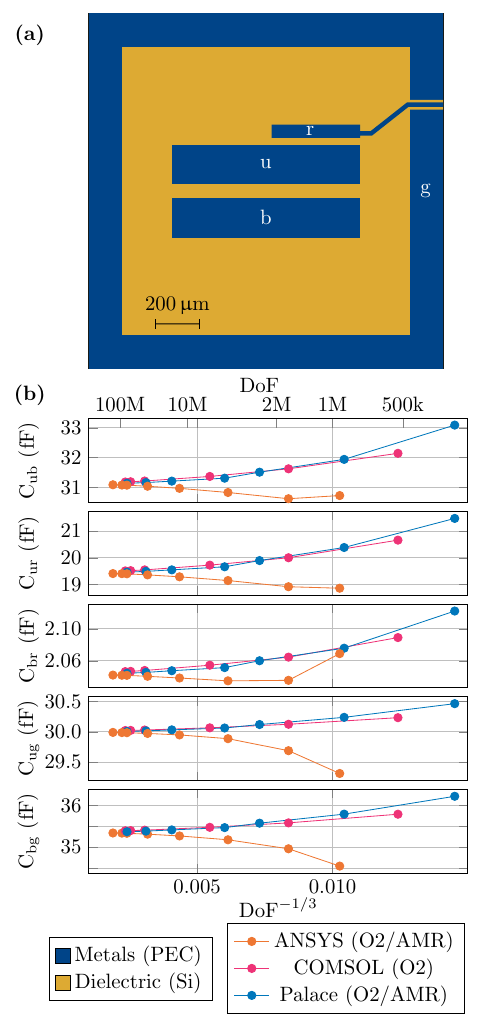}
\caption{\label{fig:cap_results} \textbf{Capacitance matrix simulations of a transmon-resonator circuit}. The design is from Fig.~\ref{fig:eig_results}(e). The simulations were performed using ANSYS, COSMOL and Palace. In each simulation, the mesh size (${\sim}{\text{DoF}}$) was decreased to converge to the continuum limit on the LHS of the plot where $\text{DoF}\to\infty$. \textbf{(a)} Electrode labels for every contiguous metallic surface along with the \textbf{(b)} relevant entries to the capacitance matrix.}
\end{figure}

Having established agreement between solvers for eigenmode simulations, we performed electrostatic simulations to verify consistency in the extraction of capacitance values between various solvers. These simulations were conducted for the design of a transmon capacitively coupled to a resonator, shown in Fig.~\ref{fig:eig_results}e. A detailed view of the transmon and the resonator coupler is presented in Fig.~\ref{fig:cap_results}a, where the key structural elements are labeled: upper capacitor pad (u), lower capacitor pad (b), readout resonator (r), and ground plane (g). Fig.~\ref{fig:cap_results}b presents the mutual capacitances extracted from the simulations as required to compute system parameters such as the qubit-resonator couplings and charging energies~\cite{blais_circuit_2021}. These capacitances are also required when using the lumped oscillator model (LOM)~\cite{groszkowski_scqubits_2021}. For this convergence study, we compare results from three solvers: the Ansys quasi-static 3D (Q3D) solver using AMR O2, the COMSOL Multiphysics electrostatic solver using MMR O2, and the Palace electrostatic solver using AMR O2. In Fig.~\ref{fig:cap_results}b, orange, red, and dark blue points represent simulations performed with Ansys, COMSOL, and Palace, respectively, plotted as a function of the notional mesh element size.

\begin{table*}[!htb] 
\caption{\label{tab: cap_results} \textbf{Final capacitance simulation results.} Comparison of the final converged capacitance values and corresponding degrees of freedom (DoF) for each solver (Palace, COMSOL, and Ansys Q3D) from the electrostatic convergence study in Fig. \ref{fig:cap_results}.}
\begin{ruledtabular}
\begin{tabular}{ccccccc}
Solver & DoF & C$_{ub}$ (fF) & C$_{ur}$ (fF) & C$_{br}$ (fF) & C$_{ug}$ (fF) & C$_{bg}$ (fF) \\ \hline
 Palace AMR O2 & 71,442,078  & 31.145  & 19.463  & 2.044 & 30.006 & 35.370 \\
 Comsol MMR O2 & 77,464,079  & 31.187  & 19.509  & 2.046 & 30.030 & 35.393 \\
 Ansys AMR O2 & 148,051,912  & 31.095  & 19.410  & 2.042 & 29.990 & 35.341 \\ 
\end{tabular}
\end{ruledtabular}
\end{table*}

As before, percentage differences from Palace (AMR O2) are shown in parentheses after the final values for the commercial solvers. The top panel in Fig.~\ref{fig:cap_results}b presents the mutual capacitance between the upper and lower transmon pads, $C_{ub}$. At the highest DoF, the Palace solver produced 31.145 fF, while COMSOL and Ansys arrived at 31.187 fF (0.135\%) and 31.095 fF (0.161\%), respectively. The second panel shows the capacitance between the upper pad and the resonator, $C_{ur}$. The final values were 19.463 fF for Palace, 19.509 fF (0.236\%) for COMSOL, and 19.410 fF (0.272\%) for Ansys. For the capacitance between the lower pad and the resonator, plotted in the third panel, $C_{br}$, at the largest DoF, Palace produced 2.044 fF, with COMSOL and Ansys yielding 2.046 fF (0.098\%) and 2.042 fF (0.098\%), respectively. The upper pad–ground plane capacitance, $C_{ug}$, produced 30.006 fF for Palace, 30.030 fF (0.080\%) for COMSOL, and 29.990 fF (0.053\%) for Ansys. Finally, the bottom panel shows the capacitance between the lower pad and the ground plane, $C_{bg}$, which yielded 35.370 fF for Palace, 35.393 fF (0.065\%) for COMSOL, and 35.341 fF (0.082\%) for Ansys. The final values for all capacitances are summarized in Table~\ref{tab: cap_results}. Across all extracted capacitances, the differences between Palace and the commercial solvers remain below 0.3\%, which is excellent numerical agreement and confirms the reliability of Palace for electrostatic simulations.

\section{Experimental Results}

Having established the accuracy of the Palace simulations using the SQDMetal API, we next compare results from SQDMetal with experimental measurements to highlight the utility of SQDMetal.

\subsection{Resonators}

\begin{figure}[!htb]
\includegraphics{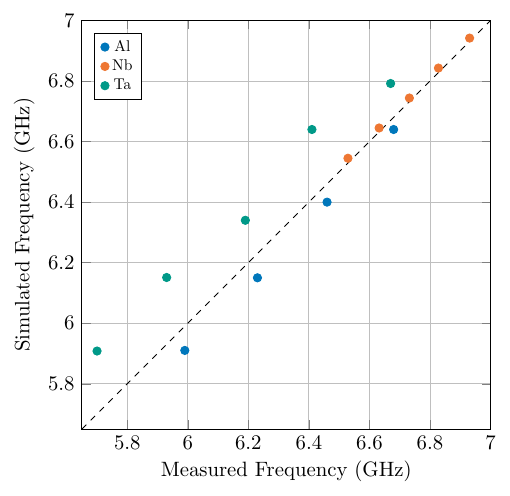}
\caption{\label{fig: res_comparison} \textbf{Comparison of simulated and measured CPW resonator frequencies.} The fundamental frequencies obtained from simulation and experiment for Al (aluminum), Nb (niobium), and Ta (tantalum) coplanar waveguide resonators are shown. Each point represents an individual resonator, with Al, Nb, and Ta denoted by blue, orange, and green circles, respectively. The dashed line indicates perfect agreement between simulated and measured values.}
\end{figure}

We begin our analysis with coplanar waveguide (CPW) resonators fabricated from various superconducting materials on a silicon substrate. Fig.~\ref{fig: res_comparison} compares the simulated and measured resonant frequencies for aluminum (Al), niobium (Nb), and tantalum (Ta) resonators. The dashed line in Fig.~\ref{fig: res_comparison} is provided as a reference, indicating perfect agreement between measured and simulated values. The simulated values used in Fig.~\ref{fig: res_comparison} were obtained using eigenmode simulations in SQDMetal (Gmsh/Palace) with fourth-order basis functions and a minimum mesh element size of 4 $\upmu\text{m}$. 

The Al and Ta $\lambda /4$ resonators were fabricated with a centre conductor width of 9 $\upmu\text{m}$ and a gap width of 5.07 $\upmu\text{m}$. The Nb $\lambda /4$ resonators were fabricated with a centre conductor and gap width of 10 $\upmu\text{m}$ and 6 $\upmu\text{m}$, respectively. These dimensions were used to ensure a characteristic impedance of $\sim$50 $\Omega$. The thickness of the Al film was 100 nm, while the Ta and Nb films were both 200 nm thick with the Ta resonators possessing approximately 100 nm of trenching. The experimental data was collected by cooling the resonator devices in a Bluefors dilution refrigerator to approximately 20 mK and measuring the resonant frequencies at high power using a vector network analyzer (VNA). We attribute any systematic offsets or deviations to be likely due to fabrication tolerances or omitted material effects.

A detailed comparison is summarized in Table~\ref{tab: res_exp_results}, which also lists the percentage difference between simulated and measured frequencies. We observe small percentage differences in Table~\ref{tab: res_exp_results}, however even closer correspondence is expected when factors such as superconductor thickness, substrate trench depth, and kinetic inductance are incorporated into simulations. Addressing these effects will be the focus of future work.

\begin{table}[!htb] 
\caption{\textbf{Summary of simulated and measured CPW resonator frequencies.} This table summarizes the data in Fig. \ref{fig: res_comparison} for the comparison of experimental and simulated values for Aluminum (Al), Niobium (Nb) and Tantalum (Ta) resonators. Additionally, the percentage difference between simulation and experiment is provided for reference.}
\label{tab: res_exp_results}
\begin{ruledtabular}
\begin{tabular}{cccc}
Resonator &Simulated (GHz) &Measured (GHz) &Difference (\%)  \\
\hline
\multicolumn{4}{c}{Aluminum Resonators}\\
\hline
1    & 5.91   & 5.99   & 1.34 \\
2    & 6.15   &6.23    & 1.28  \\
3    & 6.40   & 6.46   & 0.93   \\
4    & 6.64   & 6.68   & 0.60   \\ 
\hline
\multicolumn{4}{c}{Niobium Resonators}\\
\hline
1    & 6.545  & 6.529  & 0.24  \\
2    & 6.645   & 6.632  & 0.20   \\
3    & 6.744   & 6.732  & 0.19   \\
4    & 6.843   & 6.828  & 0.22    \\
5    & 6.942   & 6.931  & 0.16 \\
\hline
\multicolumn{4}{c}{Tantalum Resonators}\\
\hline
1    & 5.91  & 5.70  & 3.52  \\
2    & 6.15   & 5.93  & 3.59   \\
3    & 6.40   & 6.19  & 2.37   \\
4    & 6.64   & 6.41  & 3.46    \\
5    & 6.79   & 6.67  & 1.80  
\end{tabular}
\end{ruledtabular}
\end{table}

\subsection{Qubits}

After validating SQDMetal’s accuracy for resonator frequencies, we now compare its predictions for transmon qubit parameters against experimental measurements. In this comparison, we examine nine transmon qubits from two separate devices. The first device, labeled Device A, fabricated by \citet{gaikwad_entanglement_2024}, contains three transmon qubits. The second device, Device B, consists of six transmon qubits fabricated by MIT Lincoln Laboratory and was used in the work by \citet{shanto_squadds_2024}. For each qubit in both devices, we compared the simulated and measured values for the qubit transition frequency $f_{ge}$, qubit anharmonicity $\alpha_q/2\pi$, readout resonator frequency $f_r$, and coupling strength between the qubit and readout resonator $g/2\pi$. These results are summarized in Table~\ref{tab: q_results}.

\begin{table*}[!htb] 
\caption{ \label{tab: q_results} \textbf{Comparison of simulated and experimental transmon parameters.} This table lists the qubit frequency ($f_{ge}$), qubit anharmonicity ($\alpha_q$), resonator frequency ($f_r$), and coupling strength ($g$) obtained from simulation and experiment for each device. Simulated values are shown to the left of the measured values, with measured data given in square brackets. Qubits from Device A \cite{gaikwad_entanglement_2024} and Device B \cite{shanto_squadds_2024} are labeled with the prefixes A and B, respectively.}
\begin{ruledtabular}
\begin{tabular}{ccccc}
 Device &  $f_{ge}$ (GHz) & $\alpha_q/2\pi$ (MHz) &  $f_r$ (GHz) & $g/2\pi$ (MHz) \\ \hline
 AQ1& 4.60 [4.20] & 193 [212] & 7.12 [6.94]& 56 [60] \\
 AQ2& 5.13 [4.65] & 171 [180] & 7.27 [7.09] & 55 [61]\\
 AQ3& 5.61 [5.37] & 178 [140] & 7.43 [7.21] & 48 [54] \\ 
 BQ1 & 4.18 [4.22] & 144 [153] & 6.34 [6.12] & 75 [60]\\
 BQ2& 3.96 [3.90] & 142 [154] & 6.60 [6.35] & 90 [66]\\
  BQ3& 4.33 [4.45] & 173 [189] & 6.72 [6.47] & 88 [70]\\
 BQ4& 3.42 [3.59]  & 143 [164]  & 6.82 [6.57]  & 89 [66] \\
  BQ5\footnote{Due to issues with the Qiskit Metal design file qubit BQ6 was not simulated.} & 3.85 [4.10]  & 178 [210]  & 6.96 [6.66]  & 66 [52]  \\
\end{tabular}
\end{ruledtabular}
\end{table*}

We used the energy participation ratio (EPR) method was to compute the qubit parameters~\cite{minev_energy-participation_2021}. The EPR approach offers several advantages over other methods, namely, it requires only a single eigenmode simulation and it achieves good accuracy by using Maxwell’s curl equations to perform full-wave field calculations. Full-wave simulations are advantageous as they capture high-frequency effects, such as equivalent series inductance (ESL), which is often neglected in simplified models.

For Device B, the Josephson energy $E_J$ was extracted from experimental measurements of the qubit frequency and anharmonicity via~\cite{blais_circuit_2021}:

\begin{align} \label{eq:q_freq}
\hbar \omega_q = \sqrt{8E_J E_C} - E_C,
\end{align}
where $\omega_q = 2 \pi f_{ge}$ is the qubit angular frequency and $E_C$ is the charging energy, which is approximately equal to qubit anharmonicity in the transmon regime ($E_J \gg E_C$)~\cite{koch_charge-insensitive_2007}.

For Device A, applying Eqn.~\ref{eq:q_freq} directly to the measured values of $\omega_q$ and $\alpha_q$ produced $E_J$ values which were overestimated, leading to inaccurate simulation results. To improve agreement, $E_J$ was recalculated using Eqn.~\ref{eq:q_freq} and the simulated values reported in~\cite{gaikwad_entanglement_2024}. This adjustment illustrates the EPR method’s sensitivity to the choice of $E_J$ and highlights the importance of accurate experimental extraction of $E_J$ to produce consistent simulation results~\cite{minev_energy-participation_2021,yuan_comparison_2022}. Once an estimate for $E_J$ is obtained, then the corresponding Josephson inductance is calculated from:

\begin{align} \label{eq:Lj}
L_J = \frac{\phi_0^2}{E_J},
\end{align}
where $\phi_0 = \hbar/2e$ is the reduced flux quantum. For weakly nonlinear circuits operated in the dispersive regime, $|\omega_r - \omega_q| = \Delta\gg g$, such as a transmon capacitively coupled to a resonator, the Hamiltonian parameters can be derived directly from the EPRs~\cite{minev_energy-participation_2021}. The anharmonicity of the qubit and resonator, along with the dispersive shift between them, can be expressed in terms of the participation ratios as follows:

\begin{align} \alpha_q &= p_q^2 \frac{\hbar \omega_q^2}{8E_J} \label{eq: alpha_q}\\ 
\alpha_r &= p_r^2 \frac{\hbar \omega_r^2}{8E_J} \label{eq: alpha_r}\\ 
\chi_{qr} &= p_q p_r\frac{\hbar \omega_q \omega_r}{4E_J} \label{eq: chi1}, 
\end{align}
where $p_q$ and $p_r$ are the participation ratios of the qubit and resonator modes, respectively, and $\omega_q$ and $\omega_r$ are their corresponding angular frequencies. These expressions are used in this study to compute the qubit parameters in Table~\ref{tab: q_results}. However, these expressions are only appropriate for qubits operating in both the dispersive and transmon regimes. For qubits operating outside these regimes numerical diagonalization, as outlined in Appendix~\ref{EPR_section}, should be used. For reference, an explicit example of the calculation of transmon parameters using Eqns.~\ref{eq: alpha_q}, \ref{eq: alpha_r} and \ref{eq: chi1} is provided in Appendix~\ref{Transmon_Ex}.

Once the dispersive shift, $\chi_{qr}$, is determined, the coupling strength $g$ can be obtained from the following expression~\cite{shanto_squadds_2024}:

\begin{align} \label{eq: chi}
    \chi_{qr} = 2g^2 \left( \frac{\alpha_q}{\Delta(\Delta - \alpha_q)} + \frac{\alpha_q}{\Sigma(\Sigma + \alpha_q)}  \right),
\end{align}
where $\Delta = \omega_r - \omega_q$ and $\Sigma = \omega_r + \omega_q$. Eqn.~\ref{eq: chi} is proposed to be more accurate for calculating $\chi_{qr}$ because it is not derived using the rotating wave approximation (RWA), which can lead to inaccuracies due to ignoring the faster rotating terms~\cite{shanto_squadds_2024}.

To evaluate the accuracy of the simulated results, we calculated the root-mean-square error (RMSE) for each qubit parameter, both in absolute terms and as a percentage, using the data presented in Table~\ref{tab: q_results}. The resulting values, summarized in Table~\ref{tab: rmse}, show reasonable agreement between simulation and experiment. The percentage RMSE values for the qubit and resonator frequencies are 5.99\% and 3.54\%, respectively, while the anharmonicity and coupling strength exhibit larger deviations of 13.25\% and 24.53\%. These discrepancies likely stem from simplified material and geometric assumptions in the simulations, as well as from the assumption that device fabrication was performed with minimal imperfections. In particular, the coupling strength $g$ is highly sensitive to fabrication variations such as etch bias~\cite{levenson-falk_review_2024}.

The differences observed for $\alpha_q$ and $g$ in this study are consistent with values reported in other studies using the EPR method. For example, \citet{gaikwad_entanglement_2024}, who fabricated Device A, presented simulated qubit parameters from which percentage RMSE values of 17.21\% and 16.40\% were determined for $\alpha_q$ and $g$, respectively. Similarly, \citet{yuan_comparison_2022} found a 13.5\% difference for $\alpha_q$ when using the EPR method. Improved agreement between simulation and experiment is expected by incorporating additional physical effects, such as kinetic inductance~\cite{park_improving_2024}, and by modeling the full 3D device geometry, which more accurately captures fringing fields and current distributions.

Another factor contributing to the discrepancy is the omission of the Josephson junction capacitance $C_J$ in the simulations~\cite{minev_energy-participation_2021}. $C_J$ depends on the junction area and can be approximated as $C_J = 50 \pm 12$ fF/$\upmu\text{m}^2$~\cite{yilmaz_energy_2024}, which for typical Josephson junction dimensions of 260 $\times$ 180 nm results in a value of $2.3\pm0.6$ fF. This estimate is consistent with the value of 4 fF reported by \citet{minev_energy-participation_2021}. Thus, by incorporating all the aforementioned physical effects into future work on simulations, we expect greater agreement between simulated and measured qubit parameters, which will facilitate better design of superconducting qubits.

\begin{table}[!htb]
\caption{\label{tab: rmse} \textbf{Root mean square error for qubit parameters.} The RMSE is presented in absolute terms and as a percentage. The values in this table are calculated from the results in Table \ref{tab: q_results}. }
\begin{ruledtabular}
\begin{tabular}{cccc}
Parameter & Absolute RMSE (MHz) & Percentage RMSE (\%)     \\
\hline
$f_{ge}$      & 264.34  &  5.99     \\
$\alpha$      & 21.886  &  13.25    \\
$g$           & 15.564  &  24.53    \\
$f_r$         & 234.28  &  3.54     \\
\end{tabular}
\end{ruledtabular}
\end{table}

\section{Conclusion}

In this work, we have introduced SQDMetal, an open-source and highly parallel simulation workflow for superconducting quantum circuits. We demonstrated its accuracy through mesh convergence studies and benchmarking against COMSOL Multiphysics and Ansys for eigenmode and electrostatic simulations, respectively. The convergence results showed excellent agreement between all solvers, providing users of SQDMetal with a high degree of confidence in the accuracy of their simulations. We further validated SQDMetal by comparing simulated and experimental results for superconducting resonators and transmon qubits, observing reasonable agreement across qubit device parameters. Future work that incorporates material effects, such as kinetic inductance, and fully three-dimensional device geometries is expected to further improve the agreement between simulation and experiment. Overall, this study establishes SQDMetal as a reliable, community-driven, and scalable open-source platform for the accurate electromagnetic simulation and design of superconducting quantum devices.\\

\section{Acknowledgments}

We would like to acknowledge useful discussions with Eli Levenson-Falk, Saikat Das and Sadman Ahmed Shanto from the Levenson-Falk Lab (LFL) at the University of Southern California. Moreover, we would also like to thank the LFL for providing measurement data and Qiskit Metal designs for superconducting qubits, allowing the completion of the experimental results section of the paper. In addition, we would also like to thank Ahbishek Chakraborty from Chapman University who organised the collaboration with LFL and provided useful feedback on SQDMetal. We would like to acknowledge and thank Marlies Hankel and David Green from the research computing centre (RCC) at the University of Queensland (UQ). They provided valuable assistance, helping to install and run Palace on the Bunya cluster at UQ. Finally, we would like to thank Hugh Carson and Andrew Keller from AWS Palace who provided useful feedback on the mesh convergence study and provided helpful insight into Palace.\\

\bibliography{referencesActual}

\appendix

\begin{figure*} [!htb] 
\includegraphics[width=17.3cm]{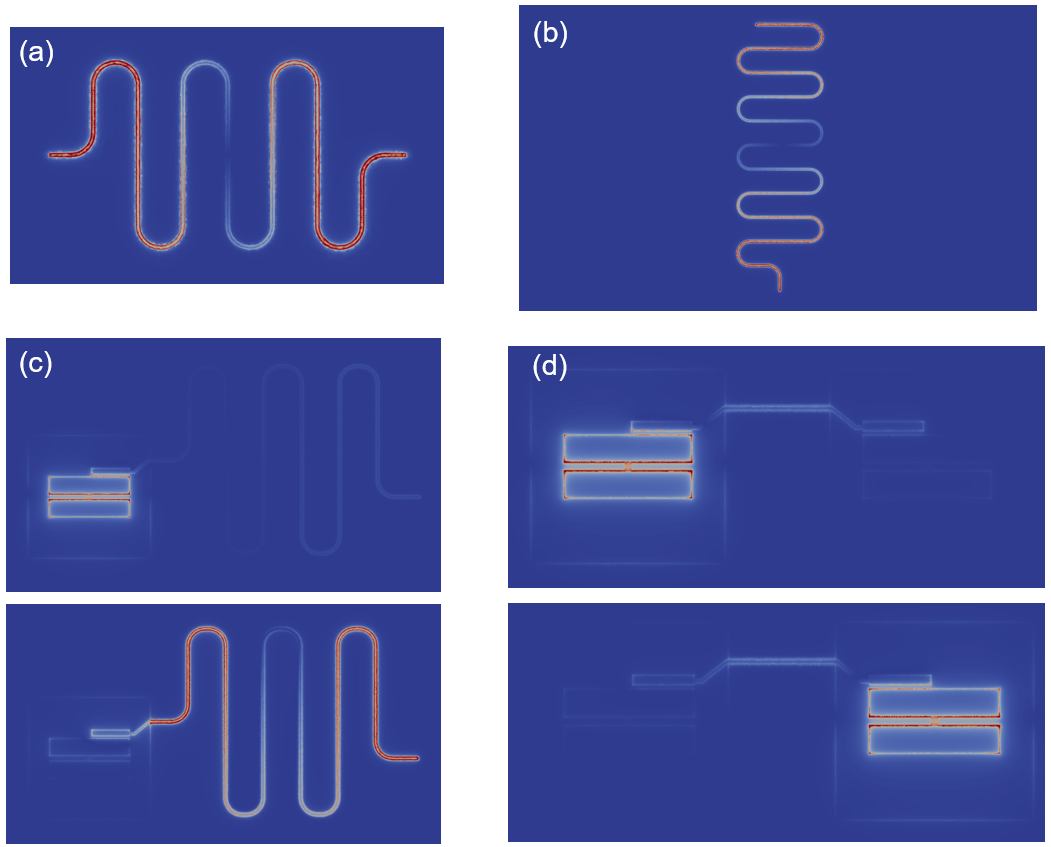}
\caption{\label{fig:results_fields} \textbf{Electric field visualization of the fundamental modes for each design.} ParaView is used to visualize the results of the eigenmode simulation. (a) Single CPW resonator. (b) Single CPW resonator capacitively coupled to a feedline. (c) Transmon capacitively coupled to a CPW resonator with the top and bottom figures showing the transmon and resonator qubit modes. (d) Two capacitively coupled transmons with the top and bottom figures showing the left and right transmon modes. }
\end{figure*}

\section{Electric Field Visualization} \label{fields_vis}

SQDMetal incorporates the use of ParaView to visualize the results of simulations. Numerous quantities of interest including the electric field, magnetic field, current density and electric energy density can be visualized with ParaView. The plots in Fig. \ref{fig:results_fields} show the electric fields for the resonant modes of each of the benchmark designs in Fig. \ref{fig:eig_results}.

\section{Energy Participation Ratio Method} \label{EPR_section}

A superconducting quantum circuit can be described as a largely linear electromagnetic environment containing one or more embedded nonlinear elements. In most cases, these nonlinear elements are Josephson junctions, although other components such as superconducting nanowires (nanoinductors) \cite{annunziata_tunable_2010} may also be used. The core idea of the energy participation ratio (EPR) method is to separate each nonlinear element into its linear and nonlinear constituents, a process known as linearization, and then remove the nonlinear part from the remaining linear circuit. An eigenmode simulation of this linearized structure is then performed to obtain the electromagnetic field distributions. From these results, the EPR is computed, quantifying the fraction of the total electromagnetic energy stored in each nonlinear element. This quantity links the linear eigenmodes of the circuit to its nonlinear dynamics and provides a direct route to determine the contribution of each nonlinear element to the system Hamiltonian. The participation ratio, which serves as the bridge between linear and nonlinear circuit components, will be discussed in greater detail in the following sections. \\

\subsection{Josephson Junction}

The energy contained in a Josephson junction is given by the following expression:

\begin{align}
    \mathcal{E}(\Phi)&= - E_J \cos{\left( \frac{\Phi}{\phi_0} \right)} ,\label{JJ_energy}
\end{align}

where $\Phi$ is the magnetic flux across the Josephson junction, $\phi_0 = \hbar/2e$ is the reduced flux quantum, and $E_J$ is a parameter known as the Josephson energy, where $E_J = I_c \phi_0$. Here, $I_c$ is the critical current of the junction and represents the maximum amount of current that can tunnel across the junction before superconductivity is broken. We can perform the Taylor expansion of Eqn. \ref{JJ_energy} as follows: 

\begin{align} \label{eq: en_expansion}
    \mathcal{E}(\Phi)&= - E_J + \frac{E_J}{2} \left( \frac{\Phi}{\phi_0} \right)^2 - \frac{E_J}{24} \left( \frac{\Phi}{\phi_0} \right)^4 + \mathcal{O}  \Big( \Phi^6 \Big), 
\end{align}

where $\mathcal{O}  ( \Phi^6 )$ represents the higher order terms in the expansion. The constant term $-E_J$ can be dropped as we are only interested in differences in potential energy not shifts. Next, we can identify the second order term as being akin to the potential energy of a linear inductor (i.e. $\Phi^2 / 2L$) with a corresponding inductance known as the Josephson inductance:

\begin{align}
    L_J = \frac{\Phi^2_0}{E_J}.
\end{align}

Then using Eqn. \ref{eq: en_expansion} we can separate out the energy of a Josephson junction into a linear and nonlinear component as follows:

\begin{align}
    \mathcal{E}(\Phi) &= \mathcal{E}(\Phi)^{\text{lin}} + \mathcal{E}(\Phi)^{\text{nlin}}\\
            &= \underbrace{\frac{\Phi^2}{2L_J}}_{\text{linear}} \underbrace{ - \frac{E_J}{24}  \left( \frac{\Phi}{\phi_0} \right)^4 + \mathcal{O}  \Big( \Phi^6 \Big)}_{\text{nonlinear}},
\end{align}

 Following this separation, the linear part of the Josephson junction can be represented in our circuit as a lumped port inductor with linear inductance, $L_J$. The nonlinear part is essentially removed from the circuit and then accounted for through the participation ratio after an eigenmode simulation of the linear circuit has been performed. As mentioned, this process of removing the nonlinearity from the qubit is referred to as linearization of the Josephson junction. 

\subsection{Full Hamiltonian of the System}

Now that the Josephson junction has been separated into its linear and nonlinear components, the full Hamiltonian of a superconducting circuit can be expressed in such a way that the linear circuit and the nonlinear circuit can be defined separately. The Hamiltonian of a linear distributed system describing an electromagnetic environment is given by:

\begin{align}
    \Hat{H}_{\textbf{lin}} = \sum_{m=1}^{M} \hbar \omega_m \Hat{a}_m^{\dag} \Hat{a}_m \label{eq:linHam}
\end{align}

where $m$ represents the electromagnetic mode, $\omega_m$ are the resonant frequencies of the modes and $\Hat{a}_m^{\dag}$ and $\Hat{a}_m$ are the creation and annihilation operators for the given mode, respectively. If we are using linear elements in our circuit we can always describe the total energy with the Hamiltonian in Eqn. \ref{eq:linHam}. Essentially, for every radiation field mode in the circuit we are associating with it a harmonic oscillator. Note that we have dropped the zero-point energy term, $\hbar \omega_m / 2$, from Eqn \ref{eq:linHam} as is customary in renormalization to avoid infinite energies. To express the full Hamiltonian we combine the linear and nonlinear parts as follows:

\begin{align}
    \Hat{H}_{\text{full}} = \sum_{m=1}^{M} \hbar \omega_m \Hat{a}_m^{\dag} \Hat{a}_m + \sum_{j=1}^{J} \mathcal{E}_j(\Hat{\Phi}_j)^{\text{nlin}}
\end{align}

where the first term, introduced in Eqn. \ref{eq:linHam}, includes the linearized Josephson junction. The second term represents the nonlinear contributions from the total number of Josephson junctions, $J$, present in the circuit. The nonlinear part of the full Hamiltonian can be expressed as:  

\begin{align}
  \mathcal{E}_j(\Hat{\phi}_j)^{\text{nlin}} &= - \frac{E_J}{24}  \Hat{\phi_j}^4 + \frac{E_J}{720}  \Hat{\phi}_j^6 - ...\\
  &= -E_J \Bigg( \cos{\left( \ \Hat{\phi}_j \right)} + \frac{1}{2} \Hat{\phi}_j^2 \Bigg),
\end{align}  

where we have now introduced the phase operator $\Hat{\phi}_j = \Hat{\Phi}_j / \phi_0$, which is known as the phase difference across the Josephson junction. We can express $\Hat{\phi}_j$ in terms of the creation and annihilation operators as follows:

\begin{align}
    \Hat{\phi}_j &= \phi^{\text{ZPF}}_{\text{m=1}}\left( \Hat{a}^{\dagger}_1 + \Hat{a}_1 \right) + \phi^{\text{ZPF}}_{\text{m=2}}\left( \Hat{a}^{\dagger}_2 + \Hat{a}_2 \right) + ... \notag \\
    &= \sum_{m=1}^{M} \phi^{\text{ZPF}}_{\text{m}}\left( \Hat{a}^{\dagger}_m + \Hat{a}_m \right), 
\end{align}

This expression shows that the phase operator, $\Hat{\Phi}$, for a given junction, $j$, is a contribution from all the modes, $m$, of the system. Now, the only unknown quantities of the full Hamiltonian are the zero point fluctuations of the phase, $\phi^{\text{ZPF}}_m$, for each mode considered in the simulation. Therefore, to describe the full Hamiltonian of the system we just need to work out $\phi^{\text{ZPF}}_m$ and this can be done using the energy participation ratio, which is defined as:

\begin{align}
    p_{mj} &= \frac{\text{Inductive energy stored in the junction, j}}{\text{Total inductive energy in mode, m}}, \label{eq: epr} \\
    &=  \frac{\langle \frac{1}{2} L_J^{-1} \Hat{\Phi}_j^2 \rangle}{\langle \frac{1}{2} \Hat{H}_{\text{lin}} \rangle},
\end{align}

where the angular brackets indicate to take the expectation value of the expression. Note that due to the energy balance between the inductive energy and capacitive energy stored in a given mode, the total inductive energy of the mode can be expressed as half the total energy. Then performing the calculation, where we take the expectation value with reference to a given Fock state $|n\rangle$, we end up with the result:

\begin{align}
    p_{mj} = \frac{2 E_J \left( \phi_{mj}^{\text{ZPF}} \right)^2 }{ \hbar \omega_m}, \label{part_ratio}
\end{align}

then rearranging to solve for $\phi_{mj}^{\text{ZPF}}$ we have

\begin{align}
    \phi_{mj}^{\text{ZPF}} = S_{mj} \sqrt{\frac{p_{mj}  \hbar \omega_m}{2 E_J} }, \label{phi_zpf}
\end{align}

where $S_{mj} \in \{-1,1\}$ is the sign of the junction and corresponds to the direction of current flow across the junction. From Eqn. \ref{phi_zpf} it can be seen that the participation ratio links the quantum properties of the circuit to its classical electromagnetic behaviour. Furthermore, all the components of the Hamiltonian are now fully specified and the system can now be solved by numerical diagonalization. The explanation provided in this section is a brief summary of the work in Ref. \cite{minev_energy-participation_2021}, for further details please refer to this paper.\\

 \section{Example: Transmon Coupled to a Resonator} \label{Transmon_Ex}

In order to demonstrate how the Hamiltonian is calculated we utilize the benchmarking example of a transmon capacitively coupled to a resonator. In this design, a Josephson inductance of 11 nH was assigned to the rectangular region between the capacitor pads, which models the Josephson junction as a lumped port. After assigning the lumped port inductance, an eigenmode simulation is run, returning the resonant modes of the system.

\begin{table}[!htb] 
\caption{\label{tab: hamiltonian} \textbf{Hamiltonian parameters for the coupled transmon and resonator system.}}
\begin{ruledtabular}
\begin{tabular}{cccc}
Parameter & Value     \\
\hline
Qubit Anharmonicity (MHz)    & 319.91   \\
Dispersive Shift (MHz)    & 2.74     \\
Qubit Frequency (GHz)    & 5.90     \\
$g$ (MHz)   & 220.06   
\end{tabular}
\end{ruledtabular}
\end{table}

The energy participation ratios for each mode are automatically calculated in post-processing by Palace. Recall that the EPR for a given mode is defined as the inductive energy stored in the junction divided by the total inductive energy in the mode as given in Eqn. \eqref{eq: epr}. Additionally, the energy balance for a resonantly excited mode, $m$, due to time-averaging is equally split between inductive energy and electric energy \cite{minev_energy-participation_2021}:

\begin{align}
    \mathcal{E}^{elec}_m = \mathcal{E}^{ind}_m, \label{e_balance}
\end{align}

where $\mathcal{E}^{elec}_m$ is the total electric energy  and $\mathcal{E}^{ind}_m$ is the total inductive stored in mode $m$ respectively. Therefore, the participation ratio for a given mode $m$ and junction $j$ is calculated by Palace as:

\begin{align}
    p_{m,j} = \frac{1}{{\mathcal{E}}^{elec}_m} \frac{1}{2} L_J |I_{m,j}|^2,
\end{align}

where $I_{m,j}$ is the peak current across junction $j$ given mode $m$ is excited.\\

For the excited transmon mode in the top panel of Fig. \ref{fig:results_fields}(c) a linearized frequency of 6.217 GHz ($\omega_q$ = 39.063 GHz) was obtained and the participation ratio is 0.99195. For the resonator mode in the bottom panel of Fig. \ref{fig:results_fields}(c) a frequency of 9.503 GHz ($\omega_r$ = 59.709 GHz) was found along with a participation ratio of 0.00278. Then, using Eqns. \eqref{eq: alpha_q}, \eqref{eq: alpha_r}, \eqref{eq: chi1} and \eqref{eq: chi} we calculate the Hamiltonian parameters as presented in Table \ref{tab: hamiltonian}. 

\section{Computers Used in the Simulations} \label{app: hardware}

\begin{figure}[!h]
\includegraphics{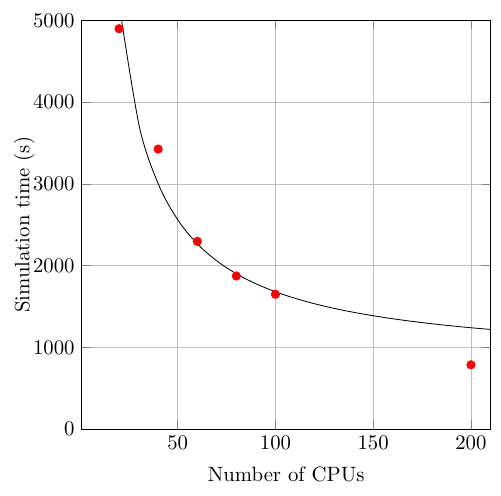}
\caption{\label{fig:bunyatimes} \textbf{Simulation times when running Palace on a cluster}. We benchmark the simulations times for an eigenmode simulation on a simple 3-resonator design with 24.5 million DoF. We fitted Amdahl's law and found a fractional worker execution efficiency of 99.2\% and a single worker execution time of approximately 88900\,s.}
\end{figure}

All Ansys and COMSOL simulations were run on a local \emph{HP DL380} computer which has a Dual \emph{Xeon E5-2697v4} CPU configuration with 768 GB of RAM. The Palace simulations were run on both this PC and on the \textit{Bunya} cluster at UQ. \textit{Bunya} possesses 113 AMD CPU nodes with 96 physical cores per compute node \cite{bunya}. The results presented in this paper used 3 nodes each with 35 CPUs (105 CPUs total) where we balanced the trade-off between execution time and the queuing allocation time. To benchmark the speed increase, we increased the number of nodes from 1 to 10 (thereby, utilizing from 20 to 200 CPUs) to observe the resulting times for an eigenmode simulation. The simulation times shown in Fig.~\ref{fig:bunyatimes} were for a triple hanging resonator design with 24.5 million DoF. We fitted Amdahl's law~\cite{Amdahl1967} to get an approximate single worker execution time of 88900\,s and a single worker execution efficiency of 99.2\%. 

\end{document}